\documentclass[11pt]{article}
\usepackage{amsmath}
\usepackage{amssymb}
\usepackage{eurosym}
\usepackage{theorem}
\usepackage{color}

\newtheorem{Lem}{Lemma}[section]
\newtheorem{Def}[Lem]{Definition}
\newtheorem{The}[Lem]{Theorem}
\newtheorem{Prop}[Lem]{Proposition}
\newtheorem{Cor}[Lem]{Corollary}

\newtheorem{Rem}[Lem]{Remark}

\begin{document}

\title{On some properties of Tsallis hypoentropies and hypodivergences}
\author{Shigeru Furuichi$^{1}$\thanks{%
E-mail:furuichi@chs.nihon-u.ac.jp},\ \ Flavia-Corina Mitroi-Symeonidis$^{2}$%
\thanks{%
E-mail:fcmitroi@yahoo.com}\ \ and Eleutherius Symeonidis$^{3}$\thanks{%
E-mail:e.symeonidis@ku-eichstaett.de} \\
$^{1}${\small Department of Information Science,}\\
{\small College of Humanities and Sciences, Nihon University,}\\
{\small 3-25-40, Sakurajyousui, Setagaya-ku, Tokyo, 156-8550, Japan}\\
$^{2}${\small Faculty of Engineering Sciences, LUMINA - University of
South-East Europe, }\\
{\small \c{S}os. Colentina 64b, Bucharest, RO-021187, Romania}\\
$^{3}${\small Mathematisch-Geographische Fakult\"{a}t, }\\
{\small Katholische Universit\"{a}t Eichst\"{a}tt-Ingolstadt, }\\
{\small \ 85071 Eichst\"{a}tt, Germany}\\
}
\date{}
\maketitle

\textbf{Abstract.} Both the Kullback-Leibler and the Tsallis divergence have
a strong limitation: if the value $0$ appears in probability distributions $%
\left( p_{1},\cdots ,p_{n}\right) $ and $\left( q_{1},\cdots ,q_{n}\right) $%
, it must appear in the same positions for the sake of significance. In
order to avoid that limitation in the framework of Shannon statistics,
Ferreri introduced in 1980 the \emph{hypoentropy}: \textquotedblleft such
conditions rarely occur in practice\textquotedblright . The aim of the
present paper is to extend Ferreri's hypoentropy to the Tsallis statistics.
We introduce the Tsallis hypoentropy and the Tsallis hypodivergence and
describe their mathematical behavior. Fundamental properties like
nonnegativity, monotonicity, the chain rule and subadditivity are
established.

 \vspace{3mm}

 \vspace{%
3mm}

\section{Preliminaries}

Throughout this paper, $X$, $Y$ and $Z$ denote discrete random variables
taking on the values $\left\{ x_{1},\cdots ,x_{|X|}\right\}$, $\left\{
y_{1},\cdots ,y_{|Y|}\right\} $ and $\left\{ z_{1},\cdots ,z_{|Z|}\right\} $%
, respectively. Where $|A|$ denotes the number of the values of the discrete
random variable $A$. We also denote the discrete random variable following a
uniform distribution by $U$. We set the probabilities as $p(x_{i})\equiv
Pr(X=x_{i})$, $p(y_{j})\equiv Pr(Y=y_{j})$ and $p(z_{k})\equiv Pr(Z=z_{k})$.
If $|U|=n$, then $p(u_{k})=\frac{1}{n}$ for all $k=1,\cdots ,n$. In
addition, we denote by $p(x_{i},y_{j})=Pr(X=x_{i},Y=y_{j})$, $%
p(x_{i},y_{j},z_{k})=Pr(X=x_{i},Y=y_{j},Z=z_{k})$ the joint probabilities,
by $p(x_{i}|y_{j})=Pr(X=x_{i}|Y=y_{j})$, $%
p(x_{i}|y_{j},z_{k})=Pr(X=x_{i}|Y=y_{j},Z=z_{k})$ the conditional
probabilities and so on.

The notion of entropy was used in statistical thermodynamics by Boltzmann 
\cite{Boltz1871} in 1871 and Gibbs \cite{Gibbs1902} in 1902, in order to
quantify the diversity, uncertainty, randomness of isolated systems. Later
it was seen as a measure of\thinspace\ \textquotedblleft information, choice
and uncertainty\textquotedblright \thinspace\ in the theory of
communication, when Shannon \cite{Sha1948} defined it by 
\begin{equation}
H(X)\equiv -\sum_{i=1}^{|X|}p(x_{i})\log p(x_{i}).
\end{equation}

In what follows we consider $|X|=|Y|=|U|=n,\ $unless otherwise specified.

Making use of the concavity of the logarithmic function, one can easily
check that the equiprobable states are maximizing the entropy, that is%
\begin{equation}
H(X)\leq H(U)=\log n.
\end{equation}
The right hand side term of this inequality is known since 1928 as Hartley
entropy \cite{Har1928}.

For two random variables $X$ and $Y$ following distributions $\{p(x_{i})\}$
and $\{p(y_{i})\}$, the Kullback-Leibler \cite{Kull-Leib1946} discrimination
function (divergence or relative entropy)\footnote{%
The relative entropy is usually defined for two probability distributions $%
P=\{p_{i}\}$ and $Q=\{q_{i}\}$ as $D(P||Q)\equiv -\sum_{i=1}^{n}p_{i}\log 
\frac{q_{i}}{p_{i}}$ in the standard notation of Information theory. $%
D(P||Q) $ is often rewritten by $D(X||Y)$ for random variables $X$ and $Y$
following the distributions $P$ and $Q$. Throughout this paper, we use the
style of Eq.(\ref{relative_entropy}) for relative entropies to unify the
notation with simple descriptions.} is defined by 
\begin{equation}
D(X||Y)\equiv \sum_{i=1}^{n}p(x_{i})(\log p(x_{i})-\log
p(y_{i}))=-\sum_{i=1}^{n}p(x_{i})\log \frac{p(y_{i})}{p(x_{i})}.
\label{relative_entropy}
\end{equation}%
Here the conventions\footnote{%
The convention is often given in the following way with the definition of $%
D(X||Y)$. If there exists $i$ such that $p(x_{i})\neq 0=p(y_{i})$, then we
define $D(X||Y)\equiv +\infty $ (in this case, $D(X||Y)$ is not significant
as an information measure any longer). Otherwise, $D(X||Y)$ is defined by
Eq.(\ref{relative_entropy}) with the convention $0\cdot \log \frac{0}{0}=0$.
This fact has been mentioned in the abstract of the paper.} $a\cdot \log 
\frac{0}{a}=-\infty \,(a>0)$ and $0\cdot \log \frac{b}{0}=0\,(b\geq 0)$ are
used. In what follows, we use such conventions in the definitions of the
entropies and divergences. However we do not state them repeatedly.

It holds that 
\begin{equation}
H(U)-H(X)=D(X||U).
\end{equation}%
Moreover, the cross-entropy (or inaccuracy) 
\begin{equation}
H^{(cross)}(X,Y)\equiv -\sum_{i=1}^{n}p(x_{i})\log p(y_{i})
\end{equation}%
satisfies the identity 
\begin{equation}
D(X||Y)=H^{(cross)}(X,Y)-H(X).
\end{equation}

C. Tsallis introduced a one-parameter extension of the entropy in 1988 in 
\cite{Tsa1988}, for handling systems which appear to deviate from standard
statistical distributions. It plays an important role in the nonextensive
statistical mechanics of complex systems, being defined as 
\begin{equation}
T_{q}(X)\equiv -\sum_{i=1}^{n}p(x_{i})^{q}\ln
_{q}p(x_{i})=\sum_{i=1}^{n}p(x_{i})\ln _{q}\frac{1}{p(x_{i})}\ \qquad (q\geq
0,q\neq 1).  \label{Tsallis_entropy}
\end{equation}%
Here the $q-$logarithmic function for $x>0$ is defined by $\ln _{q}(x)\equiv 
\frac{x^{1-q}-1}{1-q},$ which converges to the usual logarithmic function $%
\log (x)$ in the limit $q\rightarrow 1$. The Tsallis divergence (relative
entropy) \cite{Tsa1998} is given by%
\begin{equation}
S_{q}(X||Y)\equiv \sum_{i=1}^{n}p(x_{i})^{q}(\ln _{q}p(x_{i})-\ln
_{q}p(y_{i}))=-\sum_{i=1}^{n}p(x_{i})\ln _{q}\frac{p(y_{i})}{p(x_{i})}.
\end{equation}%
\hfill


\section{Hypoentropy and hypodivergence}

For nonnegative real numbers $a_{i}$ and $b_{i}\ (i=1,\cdots ,n)$, we define
the generalized relative entropy (for incomplete probability distributions): 
\begin{equation}
D^{(gen)}(a_{1},\cdots ,a_{n}\mathbf{||}b_{1},\cdots ,b_{n})\equiv
\sum_{i=1}^{n}a_{i}\log \frac{a_{i}}{b_{i}}.  \label{gen_RE}
\end{equation}

Then we have the so-called \textquotedblleft $\log $-sum\textquotedblright
\thinspace\ inequality: 
\begin{equation}
\sum_{i=1}^{n}a_{i}\log \frac{a_{i}}{b_{i}}\geq \left(
\sum_{i=1}^{n}a_{i}\right) \log \frac{\sum_{i=1}^{n}a_{i}}{%
\sum_{i=1}^{n}b_{i}},
\end{equation}%
with equality if and only if $\frac{a_{i}}{b_{i}}=const.$ for all $%
i=1,\cdots ,n$.

If we impose the condition%
\begin{equation*}
\sum_{i=1}^{n}a_{i}=\sum_{i=1}^{n}b_{i}=1,
\end{equation*}%
then $D^{(gen)}(a_{1},\cdots ,a_{n}\mathbf{||}b_{1},\cdots ,b_{n})$ is just
the relative entropy, 
\begin{equation}
D(a_{1},\cdots ,a_{n}\mathbf{||}b_{1},\cdots ,b_{n})\equiv
\sum_{i=1}^{n}a_{i}\log \frac{a_{i}}{b_{i}}.
\end{equation}

We put $a_{i}=\frac{1}{\lambda }+p(x_{i})$ and $b_{i}=\frac{1}{\lambda }%
+p(y_{i})$ with $\lambda >0$ and $\sum_{i=1}^{n}p(x_{i})=%
\sum_{i=1}^{n}p(y_{i})=1,p(x_{i})\geq 0,p(y_{i})\geq 0$. Then we find that
it is equal to the \emph{hypodivergence} ($\lambda $-divergence) introduced
by Ferreri \cite{Fer1980}, 
\begin{equation}
K_{\lambda }(X||Y)\equiv \frac{1}{\lambda }\sum_{i=1}^{n}(1+\lambda
p(x_{i}))\log \frac{1+\lambda p(x_{i})}{1+\lambda p(y_{i})}.
\end{equation}%
Clearly we have 
\begin{equation}
\lim_{\lambda \rightarrow \infty }K_{\lambda }(X||Y)=D(X||Y).
\end{equation}%
Using the \textquotedblleft $\log $-sum\textquotedblright \thinspace\
inequality, we have the nonnegativity 
\begin{equation}
K_{\lambda }(X||Y)\geq 0,  \label{nonnegativity_hypodivergence}
\end{equation}%
with equality if and only if $p(x_{i})=p(y_{i})$ for all $i=1,\cdots ,n$.

The \emph{hypoentropy }at the level $\lambda $\emph{\ }($\lambda $-entropy)
was introduced in 1980 by Ferreri \cite{Fer1980} as an alternative measure
of information in the following form: 
\begin{equation}
F_{\lambda }(X)\equiv \frac{1}{\lambda }(\lambda +1)\log (\lambda +1)-\frac{1%
}{\lambda }\sum_{i=1}^{n}(1+\lambda p(x_{i}))\log (1+\lambda p(x_{i}))\ 
\end{equation}%
for $\lambda >0.$ According to Ferreri \cite{Fer1980}, the parameter $%
\lambda $ can be interpreted as a measure of the information inaccuracy of
economic forecast. For this quantity $F_{\lambda }(X)$, we have the
following fundamental relations.

\begin{Prop}
\label{Prop - hyp} For $\lambda >0$, we have the following inequalities: 
\begin{equation}
0\leq F_{\lambda }(X)\leq F_{\lambda }(U).
\end{equation}%
The equality in the first inequality holds if and only if $p(x_{j})=1$ for
some $j$ (then $p(x_{i})=0$ for all $i\neq j$). The equality in the second
inequality holds if and only if $p(x_{i})=1/n$ for all $i=1,\cdots ,n$.
\end{Prop}

\textit{Proof:} From the nonnegativity of the hypodivergence Eq.(\ref%
{nonnegativity_hypodivergence}), we get 
\begin{eqnarray}
0 &\leq &K_{\lambda }(X||U) \\
&=&\frac{1}{\lambda }\sum_{i=1}^{n}(1+\lambda p(x_{i}))\log (1+\lambda
p(x_{i}))-\frac{1}{\lambda }(n+\lambda )\log \left( 1+\frac{\lambda }{n}%
\right) .
\end{eqnarray}%
Thus we have 
\begin{equation}
-\frac{1}{\lambda }\sum_{i=1}^{n}(1+\lambda p(x_{i}))\log (1+\lambda
p(x_{i}))\leq -\frac{1}{\lambda }(n+\lambda )\log \left( 1+\frac{\lambda }{n}%
\right) .
\end{equation}%
Adding $\frac{1}{\lambda }(\lambda +1)\log (\lambda +1)$ to both sides, we
have 
\begin{equation}
F_{\lambda }(X)\leq F_{\lambda }(U),
\end{equation}%
with equality if and only if $p(x_{i})=1/n$ for all $i=1,\cdots ,n$.

For the first inequality it is sufficient to prove: 
\begin{equation}
(1+\lambda )\log (1+\lambda )-\sum_{i=1}^{n}(1+\lambda p(x_{i}))\log
(1+\lambda p(x_{i}))\geq 0.
\end{equation}%
Since $\sum_{i=1}^{n}p(x_{i})=1$, the above inequality is written as 
\begin{equation}
\sum_{i=1}^{n}\left\{ p(x_{i})(1+\lambda )\log (1+\lambda )-(1+\lambda
p(x_{i}))\log (1+\lambda p(x_{i}))\right\} \geq 0,
\end{equation}%
so that we have only to prove 
\begin{equation}
p(x_{i})(1+\lambda )\log (1+\lambda )-(1+\lambda p(x_{i}))\log (1+\lambda
p(x_{i}))\geq 0,
\end{equation}%
for any $\lambda >0$ and $0\leq p(x_{i})\leq 1$. Lemma \ref{log} below shows
this inequality and the equality condition.

\hfill \hbox{\rule{6pt}{6pt}}

\begin{Lem}
\label{log} For any $a>0$ and $0\leq x\leq 1$, we have 
\begin{equation}
x(1+a)\log (1+a)\geq (1+ax)\log (1+ax).
\end{equation}
\end{Lem}

\textit{Proof:} We set $f(x)\equiv x(1+a)\log (1+a)-(1+ax)\log (1+ax).$ For
any $a>0$ we then have $\frac{d^{2}f(x)}{dx^{2}}=\frac{-a^{2}}{1+ax}<0$ and $%
f(0)=f(1)=0$. Thus we have the inequality.

\hfill \hbox{\rule{6pt}{6pt}}

It is a known fact that $F_{\lambda }(X)$ is monotonically increasing as a
function of $\lambda $ and 
\begin{equation}
\lim_{\lambda \rightarrow \infty }F_{\lambda }(X)=H(X),  \label{lim}
\end{equation}%
whence its name. Thus the hypoentropy appears as a generalization of
Shannon's entropy. One can see that the hypoentropy also equals zero as the
entropy does, in the case of certainty (i.e., for a so-called pure state
when all probabilities vanish but one).

\bigskip It also holds that 
\begin{equation}
F_{\lambda }(U)-F_{\lambda }(X)=K_{\lambda }(X||U).
\end{equation}%
It is of some interest for the reader to look at\ the hypoentropy which
arises for equiprobable states, \ 
\begin{equation}
F_{\lambda }(U)=\left( 1+\frac{1}{\lambda }\right) \log \left( 1+\lambda
\right) -\left( 1+\frac{n}{\lambda }\right) \log \left( 1+\frac{\lambda }{n}%
\right) .  \label{Hartley_hypoentropy}
\end{equation}%
Seen as a function of two variables, $n$ and $\lambda ,$ it increases in
each variable \cite{Fer1980}. Since 
\begin{equation}
\lim_{\lambda \rightarrow \infty }F_{\lambda }(U)=\log n,
\end{equation}%
we shall call it \emph{Hartley hypoentropy}\footnote{%
Throughout the paper we add the name \emph{Hartley} to the name of
mathematical objects whenever they are considered for the uniform
distribution. In the same way we proceed with the name \emph{Tsallis} which
we add to the name of some mathematical objects which we define, to
emphasize that they are used in the framework of Tsallis statistics. This
means that we will have \emph{Tsallis} hypoentropies, \emph{Tsallis}
hypodivergences and so on.}. We have the \emph{cross-hypoentropy} 
\begin{equation}
F_{\lambda }^{(cross)}(X,Y)\equiv \left( 1+\frac{1}{\lambda }\right) \log
\left( 1+\lambda \right) -\frac{1}{\lambda }\sum_{i=1}^{n}\left( 1+\lambda
p(x_{i})\right) \log \left( 1+\lambda p(y_{i})\right) .
\label{def_cross_hypoentropy_sec2.1}
\end{equation}%
It holds%
\begin{equation}
K_{\lambda }(X||Y)=F_{\lambda }^{(cross)}(X,Y)-F_{\lambda }(X)\geq 0,
\label{cross-hypo}
\end{equation}%
therefore we have $F_{\lambda }^{(cross)}(X,Y)\geq F_{\lambda }(X).$ This
enables us to state the following lemma.

\begin{Lem}
\label{lemma}We have the following inequality 
\begin{equation}
-\sum_{i=1}^{n}\left( 1+\lambda p(x_{i})\right) \log \left( 1+\lambda
p(x_{i})\right) \leq -\sum_{i=1}^{n}\left( 1+\lambda p(x_{i})\right) \log
\left( 1+\lambda p(y_{i})\right)  \label{ineq1}
\end{equation}%
for all $\lambda >0.$
\end{Lem}

As direct consequences we have some interesting inequalities as follows.

\begin{Prop}
\label{prop2.4_sec2.2} It holds that 
\begin{equation*}
\left( 1+\frac{\lambda }{n}\right) ^{n}\geq \prod_{i=1}^{n}\left( 1+\lambda
p(y_i)\right) ,
\end{equation*}%
for all $\lambda >0.$
\end{Prop}

\textit{Proof}: From Lemma \ref{lemma}, for $X=U$ we get%
\begin{equation}
-n\log \left( 1+\frac{\lambda }{n}\right) \leq -\sum_{i=1}^{n}\log \left(
1+\lambda p(y_{i})\right)
\end{equation}%
and the conclusion follows. \hfill \hbox{\rule{6pt}{6pt}}

An upper bound for $F_{\lambda }(X)$ can be found as follows:

\begin{Prop}
\label{prop2.5_sec2.2} The following inequality holds. 
\begin{equation*}
F_{\lambda }(X)\leq \left( 1-p_{max}\right) \log \left( 1+\lambda \right) ,
\end{equation*}%
for all $\lambda >0$, where $p_{max}\equiv \max \left\{ p(x_{1}),\cdots
,p(x_{n})\right\} $.
\end{Prop}

\bigskip \textit{\ Proof}: In Lemma \ref{lemma}, if for a fixed $k$ one
takes the probability of the $k$-th component of $Y\mathbf{\ }$to be $%
p(y_{k})=1,$ then 
\begin{equation}
-\sum_{i=1}^{n}\left( 1+\lambda p(x_{i})\right) \log \left( 1+\lambda
p(x_{i})\right) \leq -\left( 1+\lambda p(x_{k})\right) \log \left( 1+\lambda
\right) .
\end{equation}%
This implies that%
\begin{eqnarray}
F_{\lambda }(X) &\leq &\left( 1+\frac{1}{\lambda }\right) \log \left(
1+\lambda \right) -\frac{1}{\lambda }\left( 1+\lambda p(x_{k})\right) \log
\left( 1+\lambda \right) \\
&=&\left( 1-p(x_{k})\right) \log \left( 1+\lambda \right) .
\end{eqnarray}%
Since $k$ is arbitrarily fixed, the conclusion follows.

\hfill \hbox{\rule{6pt}{6pt}}

\begin{Rem}
It is of interest to notice now that, for the particular case $X=U$, we have 
\begin{equation}
F_{\lambda }(U)\leq \left( 1-\frac{1}{n}\right) \log \left( 1+\lambda
\right) .  \label{ber}
\end{equation}%
We add here one more detail: the inequality (\ref{ber}) can be verified
using Bernoulli's inequality.
\end{Rem}


\section{\protect\bigskip Tsallis hypoentropy and hypodivergence}

Now we turn our attention to the Tsallis statistics. We extend the
definition of hypodivergences as follows:

\begin{Def}
The \emph{Tsallis hypodivergence} ($q$-hypodivergence, Tsallis relative
hypoentropy) is defined by 
\begin{equation}
D_{\lambda ,q}(X||Y)\equiv -\frac{1}{\lambda }\sum_{i=1}^{n}\left( 1+\lambda
p(x_{i})\right) \ln _{q}\frac{1+\lambda p(y_{i})}{1+\lambda p(x_{i})}
\label{definition_Tsallis_Hypodivergence}
\end{equation}%
for $\lambda >0$ and $q\geq 0$.
\end{Def}

Then we have the relation: 
\begin{equation}
\lim_{\lambda \rightarrow \infty }D_{\lambda ,q}(X||Y)=S_{q}(X||Y)
\end{equation}%
which is the Tsallis divergence, and 
\begin{equation}
\lim_{q\rightarrow 1}D_{\lambda ,q}(X||Y)=K_{\lambda }(X||Y)
\end{equation}%
which is the hypodivergence.

\begin{Rem}
This definition can be also obtained from the generalized Tsallis relative
entropy (for incomplete probability distributions $\{a_{1},\cdots ,a_{n}\}$
and $\{b_{1},\cdots ,b_{n}\}$) 
\begin{equation}
D_{q}^{(gen)}(a_{1},\cdots ,a_{n}\mathbf{||}b_{1},\cdots ,b_{n})\equiv
-\sum_{i=1}^{n}a_{i}\ln _{q}\frac{b_{i}}{a_{i}},  \label{gen_qRE}
\end{equation}%
by putting $a_{i}=\frac{1}{\lambda }+p(x_{i})$ and $b_{i}=\frac{1}{\lambda }%
+p(y_{i})$ for $\lambda >0.$

The generalized relative entropy (\ref{gen_RE}) and the generalized Tsallis
relative entropy (\ref{gen_qRE}) can be written as the generalized $f$%
-divergence (for incomplete probability distributions): 
\begin{equation}
D_{f}^{(gen)}(a_{1},\cdots ,a_{n}\mathbf{||}b_{1},\cdots ,b_{n})\equiv
\sum_{i=1}^{n}a_{i}f\left( \frac{b_{i}}{a_{i}}\right)
\end{equation}%
for a convex function $f\ $on $\left( 0,\infty \right) $ and $a_{i}\geq 0$, $%
b_{i}\geq 0\ (i=1,\cdots ,n)$.
\end{Rem}

By the concavity of the $q$-logarithmic function, we have the following
\textquotedblleft $\ln _{q}$-sum\textquotedblright \thinspace\ inequality 
\begin{equation}
-\sum_{i=1}^{n}a_{i}\ln _{q}\frac{b_{i}}{a_{i}}\geq -\left(
\sum_{i=1}^{n}a_{i}\right) \ln _{q}\left( \frac{\sum_{i=1}^{n}b_{i}}{%
\sum_{i=1}^{n}a_{i}}\right) ,
\end{equation}%
with equality if and only if $\frac{a_{i}}{b_{i}}=const.$ for all $%
i=1,\cdots ,n$. Using the \textquotedblleft $\ln _{q}$-sum\textquotedblright
\thinspace\ inequality, we have the nonnegativity of the Tsallis
hypodivergence: 
\begin{equation}
D_{\lambda ,q}(X||Y)\geq 0,  \label{nonnegativity_Tsallis_Hypodivergence}
\end{equation}%
with equality if and only if $p(x_{i})=p(y_{i})$ for all $i=1,\cdots ,n$.
(The equality condition comes from the equality condition of the
\textquotedblleft $\ln _{q}$-sum\textquotedblright \thinspace\ inequality
and the condition $\sum_{i=1}^{n}p(x_{i})=\sum_{i=1}^{n}p(y_{i})=1$.)

\begin{Def}
For $\lambda >0$ and $q \geq 0$, the \emph{Tsallis hypoentropy} ($q$%
-hypoentropy) is defined by 
\begin{equation}
H_{\lambda ,q}(X)\equiv \frac{h(\lambda ,q)}{\lambda }\left\{ -(1+\lambda
)\ln _{q}\frac{1}{1+\lambda }+\sum_{i=1}^{n}(1+\lambda p(x_{i}))\ln _{q}%
\frac{1}{1+\lambda p(x_{i})}\right\}
\end{equation}%
where the function $h(\lambda ,q)>0$ satisfies two conditions, 
\begin{equation}
\lim_{q\rightarrow 1}h(\lambda ,q)=1  \label{h_condition1}
\end{equation}%
and%
\begin{equation}
\lim_{\lambda \rightarrow \infty }\frac{h(\lambda ,q)}{\lambda ^{1-q}}=1.
\label{h_condition2}
\end{equation}
\end{Def}

\bigskip These conditions are equivalent to 
\begin{equation}
\lim_{q\rightarrow 1}H_{\lambda ,q}(X)=F_{\lambda }(X)=\mathrm{Hypoentropy}
\end{equation}%
and, respectively,%
\begin{equation}
\lim_{\lambda \rightarrow \infty }H_{\lambda ,q}(X)=T_{q}(X)=\mathrm{%
Tsallis\,\,entropy.}
\end{equation}

Some interesting examples are $h(\lambda ,q)=\lambda ^{1-q}$ and $h(\lambda
,q)=(1+\lambda )^{1-q}$.

\begin{Rem}
It may be remarkable to discuss the Tsallis cross-hypoentropy. The first
candidate for the definition of the Tsallis cross-hypoentropy is 
\begin{equation}
H_{\lambda ,q}^{(cross)}(X,Y)\equiv \frac{h(\lambda ,q)}{\lambda }\left\{
-(1+\lambda )\ln _{q}\frac{1}{1+\lambda }-\sum_{i=1}^{n}(1+\lambda
p(x_{i}))^{q}\ln _{q}(1+\lambda p(y_{i}))\right\}
\label{first_candidate_Tsallis_cross_hypoentropy}
\end{equation}%
which recovers the cross-hypoentropy defined in Eq.(\ref%
{def_cross_hypoentropy_sec2.1}) in the limit $q\rightarrow 1$. Then we have 
\begin{eqnarray*}
H_{\lambda ,q}^{(cross)}(X,Y)-H_{\lambda ,q}(X) &=&\frac{h(\lambda ,q)}{%
\lambda }\sum_{i=1}^{n}\left( 1+\lambda p(x_{i})\right) ^{q}\left\{ \ln
_{q}\left( 1+\lambda p(x_{i})\right) -\ln _{q}\left( 1+\lambda
p(y_{i})\right) \right\} \\
&=&-\frac{h(\lambda ,q)}{\lambda }\sum_{i=1}^{n}\left( 1+\lambda
p(x_{i})\right) \ln _{q}\frac{1+\lambda p(y_{i})}{1+\lambda p(x_{i})} \\
&=&h(\lambda ,q)D_{\lambda ,q}(X||Y)\geq 0.
\end{eqnarray*}%
The last inequality is due to the nonnegativity given in Eq.(\ref%
{nonnegativity_Tsallis_Hypodivergence}). Since $\lim_{q\rightarrow
1}h(\lambda ,q)=1$ by the definition of the Tsallis hypoentropy (see Eq.(\ref%
{h_condition1})), the above relation recovers the inequality (\ref%
{cross-hypo}) in the limit $q\rightarrow 1$.

The second candidate for the definition of the Tsallis cross-hypoentropy is 
\begin{equation}
\tilde{H}_{\lambda ,q}^{(cross)}(X,Y)\equiv \frac{h(\lambda ,q)}{\lambda }%
\left\{ -(1+\lambda )\ln _{q}\frac{1}{1+\lambda }+\sum_{i=1}^{n}(1+\lambda
p(x_{i}))\ln _{q}\frac{1}{1+\lambda p(y_{i})}\right\}
\label{second_candidate_Tsallis_cross_hypoentropy}
\end{equation}%
which also recovers the cross-hypoentropy defined in Eq.(\ref%
{def_cross_hypoentropy_sec2.1}) in the limit $q\rightarrow 1$. Then we have 
\begin{eqnarray*}
\tilde{H}_{\lambda ,q}^{(cross)}(X,Y)-H_{\lambda ,q}(X) &=&-\frac{h(\lambda
,q)}{\lambda }\sum_{i=1}^{n}\left( 1+\lambda p(x_{i})\right) \left\{ \ln _{q}%
\frac{1}{1+\lambda p(x_{i})}-\ln _{q}\frac{1}{1+\lambda p(y_{i})}\right\} \\
&=&h(\lambda ,q)\tilde{D}_{\lambda ,q}(X||Y),
\end{eqnarray*}%
where the alternative Tsallis hypodivergence has to be defined by 
\begin{equation*}
\tilde{D}_{\lambda ,q}(X||Y)\equiv -\frac{1}{\lambda }\sum_{i=1}^{n}\left(
1+\lambda p(x_{i})\right) \left\{ \ln _{q}\frac{1}{1+\lambda p(x_{i})}-\ln
_{q}\frac{1}{1+\lambda p(y_{i})}\right\} .
\end{equation*}%
We have $\tilde{D}_{\lambda ,q}(X||Y)\neq D_{\lambda ,q}(X||Y)$ and $%
\lim_{q\rightarrow 1}\tilde{D}_{\lambda ,q}(X||Y)=K_{\lambda }(X||Y)$.
However, the nonnegativity of $\tilde{D}_{\lambda ,q}(X||Y),\,(q\geq 0)$
does not hold in general, as the following counter-examples show. Take $%
\lambda =1$, $n=2$, $p(x_{1})=0.9$, $p(y_{1})=0.8$, $q=0.5$, then $\tilde{D}%
_{\lambda ,q}(X||Y)\simeq -0.0137586.$ In addition, take $\lambda =1$, $n=3$%
, $p(x_{1})=0.3$, $p(x_{2})=0.4$, $p(y_{1})=0.2$,$p(y_{2})=0.7$ and $q=1.9$,
then $\tilde{D}_{\lambda ,q}(X||Y)\simeq -0.0195899.$ Therefore we may
conclude that Eq.(\ref{first_candidate_Tsallis_cross_hypoentropy}) is to be
given the preference over Eq.(\ref%
{second_candidate_Tsallis_cross_hypoentropy}).
\end{Rem}

We turn to show the nonnegativity and maximality for the Tsallis hypoentropy.

\begin{Lem}
\label{Lemma-q}For any $a>0$, $q \geq 0$ and $0\leq x\leq 1$, we have 
\begin{equation}
x(1+a)\ln _{q}\frac{1}{1+a}\leq (1+ax)\ln _{q}\frac{1}{1+ax}.
\end{equation}
\end{Lem}

\textit{Proof:} We set $g(x)\equiv x(1+a)\ln _{q}\frac{1}{1+a}-(1+ax)\ln _{q}%
\frac{1}{1+ax}.$ For any $a>0$ and $q \geq 0$ we then have $\frac{d^{2}g(x)}{%
dx^{2}}=qa^{2}\left( \frac{1}{1+ax}\right) ^{2-q} \geq 0$ and $g(0)=g(1)=0$.
Thus we have the inequality.

\hfill \hbox{\rule{6pt}{6pt}}

\begin{Prop}
For $\lambda >0$, $q\geq 0$ and $h(\lambda ,q)>0$ satisfying (\ref%
{h_condition1}) and (\ref{h_condition2}), we have the following
inequalities: 
\begin{equation}
0\leq H_{\lambda ,q}(X)\leq H_{\lambda ,q}(U).
\end{equation}%
The equality in the first inequality holds if and only if $p(x_{j})=1$ for
some $j$ (then $p(x_{i})=0$ for all $i\neq j$). The equality in the second
inequality holds if and only if $p(x_{i})=1/n$ for all $i=1,\cdots ,n$.
\end{Prop}

\textit{Proof:} In a similar way to the proof of Proposition \ref{Prop - hyp}%
, for the first inequality it is sufficient to prove 
\begin{equation}
-\sum_{i=1}^{n}\left\{ p(x_i)(1+\lambda )\ln _{q}\frac{1}{1+\lambda }%
-(1+\lambda p(x_i))\ln _{q}\frac{1}{1+\lambda p(x_i)}\right\} \geq 0,
\end{equation}%
so that we have only to prove 
\begin{equation}
p(x_i)(1+\lambda )\ln _{q}\frac{1}{1+\lambda }\leq (1+\lambda p(x_i))\ln _{q}%
\frac{1}{1+\lambda p(x_i)}
\end{equation}%
for any $\lambda >0$, $q \geq 0$ and $0\leq p(x_i) \leq 1$. Lemma \ref%
{Lemma-q} shows this inequality with equality condition.

The second inequality is proven by the use of the nonnegativity of the
Tsallis hypodivergence in the following way: 
\begin{equation}
0\leq D_{\lambda ,q}(X||U)=-\frac{1}{\lambda }\sum_{i=1}^{n}(1+\lambda
p(x_i))\ln _{q}\frac{1+\frac{\lambda }{n}}{1+\lambda p(x_i)}
\end{equation}%
which implies (by the use of the formula, $\ln _{q}\frac{b}{a}=b^{1-q}\ln
_{q}\frac{1}{a}+\ln _{q}b$) 
\begin{equation}
\frac{1}{\lambda }\sum_{i=1}^{n}(1+\lambda p(x_i))\ln _{q}\frac{1}{1+\lambda
p(x_i)}\leq \frac{n+\lambda }{\lambda }\ln _{q}\frac{n}{n+\lambda }.
\end{equation}%
The equality condition of the second inequality follows from the equality
condition of the nonnegativity of the Tsallis hypodivergence (\ref%
{nonnegativity_Tsallis_Hypodivergence}).

\hfill \hbox{\rule{6pt}{6pt}}

We may call 
\begin{equation*}
H_{\lambda ,q}(U)=\frac{h(\lambda ,q)}{\lambda }\left\{ -(1+\lambda )\ln _{q}%
\frac{1}{1+\lambda }+(n+\lambda )\ln _{q}\frac{1}{1+\frac{\lambda }{n}}%
\right\}
\end{equation*}%
the \emph{Hartley-Tsallis hypoentropy}. We study the monotonicity of the
Hartley-Tsallis hypoentropy $H_{\lambda ,q}(U)$ and the Tsallis hypoentropy $%
H_{\lambda ,q}(X)$.

\begin{Lem}
\label{lemma_monotone_n} The function 
\begin{equation*}
f(x)=(x+1)\ln _{q}\frac{x}{x+1}\quad (x>0)
\end{equation*}%
is monotonically increasing in $x$, for any $q\geq 0$.
\end{Lem}

\textit{Proof:} By direct calculations, we have 
\begin{equation*}
\frac{df(x)}{dx}=\frac{1}{1-q}\left\{ \left( 1+\frac{1}{x}\right)
^{q-1}\left( 1+\frac{1-q}{x}\right) -1\right\}
\end{equation*}%
and 
\begin{equation*}
\frac{d^{2}f(x)}{dx^{2}}=-qx^{-3}\left( 1+\frac{1}{x}\right) ^{q-2}\leq 0.
\end{equation*}%
Since $\lim_{x\rightarrow \infty }\frac{df(x)}{dx}=0$, we have $\frac{df(x)}{%
dx}\geq 0$.

\hfill \hbox{\rule{6pt}{6pt}}

\begin{Prop}
The Hartley-Tsallis hypoentropy 
\begin{equation*}
H_{\lambda ,q}(U)=\frac{h(\lambda ,q)}{\lambda }\left\{ -(1+\lambda )\ln _{q}%
\frac{1}{1+\lambda }+(n+\lambda )\ln _{q}\frac{1}{1+\frac{\lambda }{n}}%
\right\}
\end{equation*}%
is a monotonically increasing function of $n$, for any $\lambda >0$ and $%
q\geq 0$.
\end{Prop}

\textit{Proof:} Note that 
\begin{equation*}
H_{\lambda ,q}(U)=h(\lambda ,q)\left\{ -\left( 1+\frac{1}{\lambda }\right)
\ln _{q}\frac{1}{1+\lambda }+\left( 1+\frac{n}{\lambda }\right) \ln _{q}%
\frac{1}{1+\frac{\lambda }{n}}\right\} .
\end{equation*}%
Putting $x=\frac{n}{\lambda }>0$ for $\lambda >0$ fixed in Lemma \ref%
{lemma_monotone_n}, we get the function 
\begin{equation*}
g\left( n\right) =\left( 1+\frac{n}{\lambda }\right) \ln _{q}\frac{1}{1+%
\frac{\lambda }{n}},
\end{equation*}%
which is a monotonically increasing function of $n$. Thus we have the
present proposition.

\hfill \hbox{\rule{6pt}{6pt}}

\begin{Rem}
We have the relation 
\begin{equation*}
\lim_{n\rightarrow \infty }H_{\lambda ,q}(U)=h(\lambda ,q)\left\{ -\left( 1+%
\frac{1}{\lambda }\right) \ln _{q}\frac{1}{1+\lambda }-1\right\} .
\end{equation*}%
We notice from the condition (\ref{h_condition2}) that%
\begin{equation*}
\begin{array}{l}
\lim\limits_{\lambda \rightarrow \infty }\left( \lim\limits_{n\rightarrow
\infty }{{H_{\lambda ,q}}\left( U\right) }\right) =\lim\limits_{\lambda
\rightarrow \infty }\frac{h(\lambda ,q)}{{\lambda ^{1-q}}}\cdot {\lambda
^{1-q}\ }\left\{ {-1-\left( {1+\frac{1}{\lambda }}\right) {{\ln }_{q}}\frac{1%
}{{1+\lambda }}}\right\} \\ 
\,\,\,=\frac{1}{{1-q}}\lim\limits_{\lambda \rightarrow \infty }\frac{{{%
1+q\lambda -{{\left( {1+\lambda }\right) }^{q}}}}}{{\lambda ^{q}}}=\left\{ 
\begin{array}{l}
0\,\,\,\left( {q=0}\right) \\ 
\infty \,\,\,\,\,\left( {0<q<1}\right) \\ 
\frac{1}{{q-1}}\,\,\,\,\left( {q>1}\right) ,%
\end{array}%
\right.%
\end{array}%
\end{equation*}%
and conclude that the result is independent of the choice of $h(\lambda ,q).$

For the limit $\lambda \rightarrow 0$ we consider two cases.

\begin{itemize}
\item[(1)] In the case of $h(\lambda ,q)=\lambda ^{1-q}$, we have 
\begin{equation*}
\begin{array}{l}
\lim\limits_{\lambda \rightarrow 0}\left( \lim\limits_{n\rightarrow \infty }{%
{H_{\lambda ,q}}\left( U\right) }\right) =\lim\limits_{\lambda \rightarrow 0}%
{\lambda ^{1-q}\ }\left\{ {-1-\left( {1+\frac{1}{\lambda }}\right) {{\ln }%
_{q}}\frac{1}{{1+\lambda }}}\right\} \\ 
\,\,\,=\frac{1}{{1-q}}\lim\limits_{\lambda \rightarrow 0}\frac{{{1+q\lambda -%
{{\left( {1+\lambda }\right) }^{q}}}}}{{\lambda ^{q}}}=\left\{ 
\begin{array}{l}
\infty \,\,\,\left( {q>2}\right) \\ 
1\,\,\,\,\,\left( {q=2}\right) \\ 
0\,\,\,\,\left( {0\leq q<2}\right) ,%
\end{array}%
\right.%
\end{array}%
\end{equation*}%
as one obtains using l'H\^{o}pital's rule.

\item[(2)] In the case of $h(\lambda ,q)=(1+\lambda )^{1-q}$, we have for
all $q\geq 0$ 
\begin{equation*}
\begin{array}{l}
\lim\limits_{\lambda \rightarrow 0}\left( {\lim\limits_{n\rightarrow \infty }%
{H_{\lambda ,q}}\left( U\right) }\right) =\lim\limits_{\lambda \rightarrow 0}%
{\left( {1+\lambda }\right) ^{1-q}}\left\{ {\ -1-\left( {1+\frac{1}{\lambda }%
}\right) {{\ln }_{q}}\frac{1}{{1+\lambda }}}\right\} \\ 
=\frac{1}{{1-q}}\lim\limits_{\lambda \rightarrow 0}\frac{{{1+q\lambda -{{%
\left( {1+\lambda }\right) }^{q}}}}}{{\lambda {{\left( {1+\lambda }\right) }%
^{q-1}}}}=\frac{q}{{1-q}}\lim\limits_{\lambda \rightarrow 0}\frac{{1-{{%
\left( {1+\lambda }\right) }^{q-1}}}}{{{{\left( {1+\lambda }\right) }^{q-1}}%
+\left( {q-1}\right) \lambda {{\left( {1+\lambda }\right) }^{q-2}}}}=0.%
\end{array}%
\end{equation*}
\end{itemize}

These results mean that our Hartley-Tsallis hypoentropy with $h(\lambda
,q)=\lambda ^{1-q}$ or $(1+\lambda )^{1-q}$ has the same limits as the
Hartley hypoentropy, $F_{\lambda }(U)\ $(see also \cite{Fer1980}), in the
case $0<q<1$.
\end{Rem}

We study here the monotonicity of $H_{\lambda ,q}(X)$ for $h(\lambda
,q)=(1+\lambda )^{1-q}.$ The other case $h(\lambda ,q)=\lambda ^{1-q}$ is
studied in the next section, see Lemma \ref{lemma_monotonicity}.

\begin{Prop}
\label{Tsallis_hypoentropy_monotonicity_ineq01} We assume $h(\lambda
,q)=(1+\lambda )^{1-q}$. Then $H_{\lambda ,q}(X)$ is a monotone increasing
function of $\lambda >0$ when $0\leq q\leq 2$.
\end{Prop}

\textit{Proof:} Note that 
\begin{equation*}
H_{\lambda ,q}(X)=\sum_{i=1}^{n}Sn_{\lambda,q}(p(x_{i})),
\end{equation*}%
where 
\begin{equation*}
Sn_{\lambda,q}(x)\equiv \frac{(1+\lambda )^{1-q}}{\lambda (1-q)}\left\{
(1+\lambda x)^{q}-(1+\lambda )^{q}x+x-1\right\}
\end{equation*}%
is defined on $0\leq x\leq 1$, $0\leq q\leq 2$ and $\lambda >0$. Then we
have 
\begin{equation*}
\frac{dH_{\lambda ,q}(X)}{d\lambda }=\sum_{i=1}^{n} \frac{d Sn_{\lambda
,q}(p(x_{i}))}{d\lambda }=\sum_{i=1}^{n}s_{\lambda,q}(p(x_{i})),
\end{equation*}%
where 
\begin{equation*}
s_{\lambda ,q}(x)\equiv \frac{q\lambda (1-x)\left\{ 1-(1+\lambda
x)^{q-1}\right\} +1-x+(1+\lambda )^{q}x-(1+\lambda x)^{q}}{(1-q)\lambda
^{2}(1+\lambda )^{q}}
\end{equation*}%
is defined on $0\leq x\leq 1$, $0\leq q\leq 2$ and $\lambda >0$. By some
computations, we have 
\begin{equation*}
\frac{d^{2}s_{\lambda ,q}(x)}{dx^{2}}=\frac{-q(1+\lambda x)^{q-3}\left[
1+\lambda \left\{ (x-1)(q-1)+1\right\} \right] }{(1+\lambda )^{q}}\leq 0,
\end{equation*}%
since $(x-1)(q-1)+1\geq 0$ for $0\leq x\leq 1$ and $0\leq q\leq 2$. We
easily find $s_{\lambda ,q}(0)=s_{\lambda ,q}(1)=0$. Thus we have $%
s_{\lambda ,q}(x)\geq 0$ for $0\leq x\leq 1$, $0\leq q\leq 2$ and $\lambda
>0 $. Therefore we have $\frac{dH_{\lambda ,q}(X)}{d\lambda }\geq 0$ for $%
0\leq q\leq 2$ and $\lambda >0$.

\hfill \hbox{\rule{6pt}{6pt}}

This result agrees with the known fact that the usual (Ferreri) hypoentropy
is increasing as a function of $\lambda$.

Closing this subsection, we give a $q$-extended version for Proposition \ref%
{prop2.5_sec2.2} and Proposition \ref{prop2.4_sec2.2}.

\begin{Prop}
Let $p_{max}\equiv \max \{p(x_{1}),\cdots ,p(x_{n})\}$. Then we have the
following inequality. 
\begin{equation}
H_{\lambda ,q}(X)\leq \frac{h(\lambda ,q)}{\lambda }\left\{ (1+\lambda
)^{q}-(1+\lambda p_{max})^{q}\right\} \ln _{q}(1+\lambda )
\end{equation}%
for all $\lambda >0$ and $q\geq 0$.
\end{Prop}

\textit{Proof:} From the \thinspace \textquotedblleft $\ln _{q}$%
-sum\textquotedblright \thinspace\ inequality, we have $D_{\lambda
,q}(X||Y)\geq 0$. Since $\lambda >0$, we have 
\begin{equation}
-\sum_{i=1}^{n}(1+\lambda p(x_{i}))\ln _{q}\frac{1+\lambda p(y_{i})}{%
1+\lambda p(x_{i})}\geq 0  \label{1}
\end{equation}%
which is equivalent to$\ $ 
\begin{equation}
\sum_{i=1}^{n}(1+\lambda p(x_{i}))^{q}\left\{ \ln _{q}(1+\lambda
p(x_{i}))-\ln _{q}(1+\lambda p(y_{i}))\right\} \geq 0.  \label{2}
\end{equation}%
Thus we have 
\begin{equation}
-\sum_{i=1}^{n}(1+\lambda p(x_{i}))^{q}\ln _{q}(1+\lambda p(x_{i}))\leq
-\sum_{i=1}^{n}(1+\lambda p(x_{i}))^{q}\ln _{q}(1+\lambda p(y_{i})),
\label{ineq01_proof_prop2.14_sec2.2}
\end{equation}%
which extends the result of Lemma \ref{lemma}. For arbitrarily fixed $k$, we
set $p(y_{k})=1$ (and $p(y_{i})=0$ for $i\neq k$) in the above inequality,
then we have 
\begin{equation}
-\sum_{i=1}^{n}(1+\lambda p(x_{i}))^{q}\ln _{q}(1+\lambda p(x_{i}))\leq
-(1+\lambda p(x_{k}))^{q}\ln _{q}(1+\lambda ).
\end{equation}%
Since $x^{q}\ln _{q}x=-x\ln _{q}\frac{1}{x}$, we have 
\begin{equation}
\sum_{i=1}^{n}(1+\lambda p(x_{i}))\ln _{q}\frac{1}{1+\lambda p(x_{i})}\leq
-(1+\lambda p(x_{k}))^{q}\ln _{q}(1+\lambda ).
\end{equation}%
Multiplying both sides by $\frac{h(\lambda ,q)}{\lambda }>0$ and then adding 
\begin{equation}
-\frac{h(\lambda ,q)}{\lambda }(1+\lambda )\ln _{q}\frac{1}{1+\lambda }=%
\frac{h(\lambda ,q)}{\lambda }(1+\lambda )^{q}\ln _{q}(1+\lambda )
\end{equation}%
to both sides, we have 
\begin{equation}
H_{\lambda ,q}(X)\leq \frac{h(\lambda ,q)}{\lambda }\left\{ (1+\lambda
)^{q}-(1+\lambda p(x_{k}))^{q}\right\} \ln _{q}(1+\lambda ).
\end{equation}%
Since $k$ is arbitrary, we have this proposition.

\hfill \hbox{\rule{6pt}{6pt}}

Letting $q\rightarrow 1$ in the above proposition, we recover Proposition %
\ref{prop2.5_sec2.2}.\textit{\ }

We give some notations before we state the next proposition. For any $x,y>0$
satisfying $x^{1-q}+y^{1-q}-1>0$, we define the $q$-product \cite{Su02} by 
\begin{equation*}
x\otimes _{q}y\equiv \left( x^{1-q}+y^{1-q}-1\right) ^{\frac{1}{1-q}}.
\end{equation*}%
Then we have $\lim_{q\rightarrow 1}x\otimes _{q}y=xy$ and $\ln _{q}(x\otimes
_{q}y)=\ln _{q}x+\ln _{q}y$. We also use the notation ${x^{\otimes _{q}^{n}}}%
=\underbrace{x{\ \otimes _{q}}\cdots {\ \otimes _{q}}x}_{n}$ and $%
\mathop {{
\otimes _q}}\limits_{j=1}^{n}\left( {x_{j}}\right) ={x_{1}}{\ \otimes _{q}}%
\cdots {\ \otimes _{q}}{x_{n}}$.

\begin{Prop}
\begin{equation*}
{\left( {1+\frac{\lambda }{n}}\right) ^{\otimes _{q}^{n}}}\geq \mathop {{
\otimes _q}}\limits_{i=1}^{n}\left( {1+\lambda {p(y_i)}}\right)
\end{equation*}%
for all $\lambda >0$ and $0\leq q<1$.
\end{Prop}

\textit{Proof:} In the inequality (\ref{ineq01_proof_prop2.14_sec2.2}), we
put $p(x_i)=\frac{1}{n}$ for all $i=1,\cdots ,n$. Then we have 
\begin{equation*}
n\ln _{q}\left( 1+\frac{\lambda }{n}\right) \geq \sum_{i=1}^{n}\ln
_{q}\left( 1+\lambda p(y_i)\right) ,
\end{equation*}%
which implies this proposition.

\hfill \hbox{\rule{6pt}{6pt}}

The limit $q\rightarrow 1$ in the above proposition recovers Proposition \ref%
{prop2.4_sec2.2}. In addition, it is known that $\lim_{n\rightarrow \infty
}\left( 1+\frac{\lambda }{n}\right) ^{\otimes _{q}^{n}}=\exp _{q}(\lambda )$%
, where $\exp _{q}(x)$ is the inverse function of $\ln _{q}(x)$ and defined
as $\exp _{q}(x)\equiv \left\{ 1+(1-q)x\right\} ^{\frac{1}{1-q}}$ for the
case $1+(1-q)x>0$. 

\section{The subadditivities of the Tsallis hypoentropies}

Throughout this section we assume $|X|=n,$ $|Y|=m,|Z|=l.$ We define the\emph{%
\ joint Tsallis hypoentropy} at the level $\lambda $ by 
\begin{equation}
H_{\lambda ,q}(X,Y)\equiv \frac{h(\lambda ,q)}{\lambda }\left\{ -(1+\lambda
)\ln _{q}\frac{1}{1+\lambda }+\sum_{i=1}^{n}\sum_{j=1}^{m}(1+\lambda
p(x_{i},y_{j}))\ln _{q}\frac{1}{1+\lambda p(x_{i},y_{j})}\right\} .
\end{equation}%
Note that $H_{\lambda ,q}(X,Y)=H_{\lambda ,q}(Y,X)$.

For all $i=1,\cdots ,n$ for which $p(x_{i})\neq 0,$ we define the Tsallis
hypoentropy of $Y$ given $X=x_{i}$, at the level $\lambda p(x_{i}),$ by 
\begin{eqnarray}
&&H_{\lambda p(x_{i}),q}(Y|x_{i})  \notag \\
&\equiv &\frac{h(\lambda p(x_{i}),q)}{\lambda p(x_{i})}\left\{ -\left(
1+\lambda p(x_{i})\right) \ln _{q}\frac{1}{1+\lambda p(x_{i})}%
+\sum\limits_{j=1}^{m}\left( 1+\lambda p(x_{i}) p(y_{j}\vert x_{i}) \right)
\ln _{q}\frac{1}{1+\lambda p(x_{i}) p(y_{j}\vert x_{i})}\right\}  \notag \\
&=&\frac{h(\lambda p(x_{i}),q)}{\lambda p(x_{i})}\left\{ -\left( 1+\lambda
p(x_{i})\right) \ln _{q}\frac{1}{1+\lambda p(x_{i})}+\sum\limits_{j=1}^{m}%
\left( 1+\lambda p(x_{i},y_{j})\right) \ln _{q}\frac{1}{1+\lambda
p(x_{i},y_{j})}\right\} .
\end{eqnarray}

For $n=1$, this coincides with the hypoentropy $H_{\lambda ,q}(Y).$ As for
the particular case $m=1,$ we get $H_{\lambda p(x_{i}),q}(Y|x_{i})=0.$

\begin{Def}
The \emph{Tsallis conditional hypoentropy\textbf{\ }}at the level\emph{\ }$%
\lambda $ is defined by 
\begin{equation}
H_{\lambda ,q}(Y|X)\equiv \sum_{i=1}^{n}p(x_{i})^{q}H_{\lambda
p(x_{i}),q}(Y|x_{i}).  \label{definition_Tsallis_conditional_hypoentropy}
\end{equation}%
(As a usual convention, the corresponding summand is defined as $0$, if $%
p(x_{i})=0$. )
\end{Def}

Throughout this section we consider the particular function $h(\lambda
,q)=\lambda ^{1-q}\ $for $\lambda >0,$ $q\geq 0$.

\begin{Lem}
\label{lemma01_sec2.2} We assume $h(\lambda,q)=\lambda^{1-q}$. The chain
rule for the Tsallis hypoentropy holds: 
\begin{equation}
H_{\lambda ,q}(X,Y)=H_{\lambda ,q}(X)+H_{\lambda ,q}(Y|X).
\label{lambda_q_add}
\end{equation}
\end{Lem}

\textit{Proof:} The proof is done by straightforward computation as follows. 
\begin{equation*}
\begin{array}{l}
{H_{\lambda ,q}}\left( X\right) +{H_{\lambda ,q}}\left( {Y|X}\right) =\frac{{%
{\lambda ^{1-q}}}}{\lambda }\left\{ {-\left( {1+\lambda }\right) {{\ln }_{q}}%
\frac{1}{{1+\lambda }}+\sum\limits_{i=1}^{n}{\left( {1+\lambda p(x_i) }%
\right) {{\ln }_{q}}\frac{1}{{1+\lambda p(x_i) }}}}\right\} \\ 
\,\,\,\,\,\,\,\,+\sum\limits_{i=1}^{n}{\frac{{{{\left( {\lambda p(x_i) }%
\right) }^{1-q}}}}{{\lambda p(x_i) }}p(x_i) ^{q}\left\{ {-\left( {1+\lambda {%
p(x_i) }}\right) {{\ln }_{q}}\frac{1}{{1+\lambda p(x_i) }}%
+\sum\limits_{j=1}^{m}{\left( {1+\lambda {p(x_i,y_j) }}\right) {{\ln }_{q}}%
\frac{1}{{1+\lambda {p(x_i,y_j) }}}}}\right\} } \\ 
=\frac{{{\lambda ^{1-q}}}}{\lambda }\left\{ {-\left( {1+\lambda }\right) {{%
\ln }_{q}}\frac{1}{{1+\lambda }}+\sum\limits_{i=1}^{n}{\left( {1+\lambda {%
p(x_i) }}\right) {{\ln }_{q}}\frac{1}{{1+\lambda {p(x_i) }}}}}\right\} \\ 
\,\,\,\,\,\,\,+\frac{{{\lambda ^{1-q}}}}{\lambda }\left\{ {%
-\sum\limits_{i=1}^{n}{\left( {1+\lambda {p(x_i) }}\right) {{\ln }_{q}}\frac{%
1}{{1+\lambda {p(x_i) }}}}+\sum\limits_{i=1}^{n}{\sum\limits_{j=1}^{m}{%
\left( {1+\lambda {p(x_i,y_j) }}\right) {{\ln }_{q}}\frac{1}{{1+\lambda {%
p(x_i,y_j) }}}}}}\right\} \\ 
=\frac{{{\lambda ^{1-q}}}}{\lambda }\left\{ {-\left( {1+\lambda }\right) {{%
\ln }_{q}}\frac{1}{{1+\lambda }}+\sum\limits_{i=1}^{n}{\sum\limits_{j=1}^{m}{%
\left( {1+\lambda {p(x_i,y_j) }}\right) {{\ln }_{q}}\frac{1}{{1+\lambda {%
p(x_i,y_j) }}}}}}\right\} ={H_{\lambda ,q}}\left( {X,Y}\right) .%
\end{array}%
\end{equation*}

\hfill \hbox{\rule{6pt}{6pt}}

In the limit $\lambda \rightarrow \infty ,$ the identity (\ref{lambda_q_add}%
) becomes $T_{q}(X,Y)=T_{q}(X)+T_{q}(Y|X),$ where $T_{q}(Y|X)\equiv
\sum_{i=1}^{n}p(x_{i})^{q}T_{q}(Y|x_{i})=-\sum_{i=1}^{n}%
\sum_{j=1}^{m}p(x_{i},y_{j})^{q}\ln _{q}p(y_{j}|x_{i})$ is the \emph{Tsallis
conditional entropy} and $T_{q}(X,Y)\equiv {\sum\limits_{i=1}^{n}{%
\sum\limits_{j=1}^{m}p(x_{i},y_{j})}}\ln _{q}\frac{1}{{{p(x_{i},y_{j})}}}$
is the \emph{Tsallis joint entropy} (see also $\cite[p.3]{Fur2005}$).

In the limit $q\rightarrow 1$ in Lemma \ref{lemma01_sec2.2}, we also obtain
the identity $F_{\lambda }(X,Y)=F_{\lambda }(X)+F_{\lambda }(Y|X)$, which
naturally leads to the definition of $F_{\lambda }(Y|X)$ as \emph{%
conditional hypoentropy.}

In order to obtain the subadditivity for the Tsallis hypoentropy, we prove
the monotonicity of the Tsallis hypoentropy.

\begin{Lem}
\label{lemma_monotonicity} We assume $h(\lambda ,q)=\lambda ^{1-q}$. The
Tsallis hypoentropy $H_{\lambda ,q}(X)$ is a monotonically increasing
function of $\lambda >0$ when $0\leq q\leq 2$ and a monotonically decreasing
function of $\lambda >0$ when $q\geq 2$ (or $q\leq 0$).
\end{Lem}

\textit{Proof:} Note that 
\begin{equation}
H_{\lambda ,q}(X)=\sum_{i=1}^{n}Ln_{\lambda ,q}(p(x_{i})),
\end{equation}%
where 
\begin{equation}
Ln_{\lambda ,q}(x)\equiv \frac{(1+\lambda x)^{q}-(1+\lambda )^{q}x+x-1}{%
\lambda ^{q}(1-q)}
\end{equation}%
is defined on $0\leq x\leq 1$ and $\lambda >0$. Then we have 
\begin{equation}
\frac{dH_{\lambda ,q}(X)}{d\lambda }=\sum_{i=1}^{n}\frac{dLn_{\lambda
,q}(p(x_{i}))}{d\lambda }=\sum_{i=1}^{n}l_{\lambda ,q}(p(x_{i})),
\end{equation}%
where 
\begin{equation}
l_{\lambda ,q}(x)\equiv \frac{q}{\lambda ^{2}(1-q)}\left\{ \left( \frac{1}{%
\lambda }+1\right) ^{q-1}x-\left( \frac{1}{\lambda }+x\right) ^{q-1}-\frac{%
(x-1)}{\lambda ^{q-1}}\right\}
\end{equation}%
is defined on $0\leq x\leq 1$ and $\lambda >0$. By elementary computations,
we obtain 
\begin{equation}
\frac{d^{2}l_{\lambda ,q}(x)}{dx^{2}}=q(q-2)\lambda ^{1-q}(1+\lambda
x)^{q-3}.
\end{equation}%
Since we have $l_{\lambda ,q}(0)=l_{\lambda ,q}(1)=0$, we find that $%
l_{\lambda ,q}(x)\geq 0$ for $0\leq q\leq 2$ and any $\lambda >0$. We also
find that $l_{\lambda ,q}(x)\leq 0$ for $q\geq 2$ (or $q\leq 0$) and any $%
\lambda >0$. Therefore we have $\frac{dH_{\lambda ,q}(X)}{d\lambda }\geq 0$
when $0\leq q\leq 2$, and $\frac{dH_{\lambda ,q}(X)}{d\lambda }\leq 0$ when $%
q\geq 2$ (or $q\leq 0$).

\hfill \hbox{\rule{6pt}{6pt}}

This result also agrees with the known fact that the usual (Ferreri)
hypoentropy is increasing as a function of $\lambda$.

\begin{The}
\label{the01_sec2.2} We assume $h(\lambda,q)=\lambda^{1-q}$. It holds $%
H_{\lambda ,q}(Y|X)\leq H_{\lambda ,q}(Y)$ for $1\leq q\leq 2$.
\end{The}

\textit{Proof:} We note that $Ln_{\lambda ,q}(x)$ is a nonnegative and
concave function in $x$, when $0\leq x\leq 1$, $\lambda >0$ and $q\geq 0$.
Here we use the notation for the conditional probability as $p(y_{j}|x_{i})=%
\frac{p(x_{i},y_{j})}{p(x_{i})}$ when $p(x_{i})\neq 0$. By the concavity of $%
Ln_{\lambda ,q}(x)$, we have 
\begin{eqnarray}
\sum_{i=1}^{n}p(x_{i})Ln_{\lambda ,q}\left( p(y_{j}|x_{i})\right) &\leq
&Ln_{\lambda ,q}\left( \sum_{i=1}^{n}p(x_{i})p(y_{j}|x_{i})\right) \\
&=&Ln_{\lambda ,q}\left( \sum_{i=1}^{n}p(x_{i},y_{j})\right) =Ln_{\lambda
,q}(p(y_{j})).
\end{eqnarray}%
Summing both sides of the above inequality over $j$, we have 
\begin{equation}
\sum_{i=1}^{n}p(x_{i})\sum_{j=1}^{m}Ln_{\lambda ,q}\left(
p(y_{j}|x_{i})\right) \leq \sum_{j=1}^{m}Ln_{\lambda ,q}(p(y_{j})).
\label{the-proof-ineq01}
\end{equation}%
Since $p(x_{i})^{q}\leq p(x_{i})$ for $1\leq q\leq 2$ and $Ln_{\lambda
,q}(x)\geq 0$ for $0\leq x\leq 1$, $\lambda >0$ and $q\geq 0$, we have 
\begin{equation}
p(x_{i})^{q}\sum_{j=1}^{m}Ln_{\lambda ,q}\left( p(y_{j}|x_{i})\right) \leq
p(x_{i})\sum_{j=1}^{m}Ln_{\lambda ,q}\left( p(y_{j}|x_{i})\right) .
\end{equation}%
Summing both sides of the above inequality over $i$, we have 
\begin{equation}
\sum_{i=1}^{n}p(x_{i})^{q}\sum_{j=1}^{m}Ln_{\lambda ,q}\left(
p(y_{j}|x_{i})\right) \leq \sum_{i=1}^{n}p(x_{i})\sum_{j=1}^{m}Ln_{\lambda
,q}\left( p(y_{j}|x_{i})\right) .  \label{the-proof-ineq02}
\end{equation}%
By the two inequalities (\ref{the-proof-ineq01}) and (\ref{the-proof-ineq02}%
), we have 
\begin{equation}
\sum_{i=1}^{n}p(x_{i})^{q}\sum_{j=1}^{m}Ln_{\lambda ,q}\left(
p(y_{j}|x_{i})\right) \leq \sum_{j=1}^{m}Ln_{\lambda ,q}(p(y_{j})).
\label{the-proof-ineq03}
\end{equation}%
Here we can see that $\sum_{j=1}^{m}Ln_{\lambda ,q}\left(
p(y_{j}|x_{i})\right) $ is the Tsallis hypoentropy for fixed $x_{i}$ and the
Tsallis hypoentropy is a monotonically increasing function of $\lambda $ in
the case $1\leq q\leq 2$, due to Lemma \ref{lemma_monotonicity}. Thus we
have 
\begin{equation}
\sum_{j=1}^{m}Ln_{\lambda p(x_{i}),q}\left( p(y_{j}|x_{i})\right) \leq
\sum_{j=1}^{m}Ln_{\lambda ,q}\left( p(y_{j}|x_{i})\right) .
\label{the-proof-ineq04}
\end{equation}%
By the two inequalities (\ref{the-proof-ineq03}) and (\ref{the-proof-ineq04}%
), we finally have 
\begin{equation}
\sum_{i=1}^{n}p(x_{i})^{q}\sum_{j=1}^{m}Ln_{\lambda p(x_{i}),q}\left(
p(y_{j}|x_{i})\right) \leq \sum_{j=1}^{m}Ln_{\lambda ,q}(p(y_{j})),
\end{equation}%
which implies (since $p(y_{j}|x_{i})=\frac{p(x_{i},y_{j})}{p(x_{i})}$) 
\begin{equation}
\sum_{i=1}^{n}p(x_{i})^{q}H_{\lambda p(x_{i}),q}\left( Y|x_{i}\right) \leq
\sum_{j=1}^{m}Ln_{\lambda ,q}(p(y_{j})),
\end{equation}%
since we have for all fixed $x_i$, 
\begin{eqnarray*}
&&H_{\lambda p(x_{i}),q}(Y|x_{i})=\frac{1}{\lambda^q p(x_{i})^q}%
\sum_{j=1}^{m}\left\{ -p(y_{j}|x_{i})\left( 1+\lambda p(x_{i})\right) \ln
_{q}\frac{1}{1+\lambda p(x_{i})}\right. \\
&&\left. \hspace*{3.0cm}+(1+\lambda p(x_{i})p(y_{j}|x_{i}))\ln _{q}\frac{1}{%
1+\lambda p(x_{i})p(y_{j}|x_{i})}\right\} =\sum_{j=1}^{m}Ln_{\lambda
p(x_{i}),q}(p(y_{j}|x_{i})).
\end{eqnarray*}%
Therefore we have $H_{\lambda ,q}(Y|X)\leq H_{\lambda ,q}(Y)$. \hfill %
\hbox{\rule{6pt}{6pt}}

\begin{Cor}
\label{cor01_sec2.2} We have the following subadditivity for the Tsallis
hypoentropies: 
\begin{equation}
H_{\lambda ,q}(X,Y)\leq H_{\lambda ,q}(X)+H_{\lambda ,q}(Y)
\end{equation}%
in the case $h(\lambda,q)=\lambda^{1-q}$ for $1\leq q\leq 2$.
\end{Cor}

\textit{Proof:} The proof is easily done by Lemma \ref{lemma01_sec2.2} and
Theorem \ref{the01_sec2.2}.

\hfill \hbox{\rule{6pt}{6pt}}

We are now in a position to prove the strong subadditivity for the Tsallis
hypoentropies. The strong subadditivity for entropy is one of interesting
subjects in entropy theory \cite{PV}. For this purpose, we firstly give a
chain rule for three random variables $X,Y$ and $Z$.

\begin{Lem}
\label{lemma02_sec2.2} We assume $h(\lambda,q)=\lambda^{1-q}$. The following
chain rule holds: 
\begin{equation}
H_{\lambda ,q}(X,Y,Z)=H_{\lambda ,q}(X|Y,Z)+H_{\lambda ,q}(Y,Z).
\end{equation}
\end{Lem}

\textit{Proof:} The proof can be done following the recipe used in Lemma \ref%
{lemma01_sec2.2}. 
\begin{equation*}
\begin{array}{l}
{H_{\lambda ,q}}\left( {X|Y,Z}\right) +{H_{\lambda ,q}}\left( {Y,Z}\right)
\\ 
=\sum\limits_{j=1}^{m}{\sum\limits_{k=1}^{l}p{{{\left( {{y_{j}},{z_{k}}}%
\right) }^{q}}}}\frac{1}{{{{\left( {\lambda p\left( {{y_{j}},{z_{k}}}\right) 
}\right) }^{q}}}}\left\{ {-\left( {1+\lambda p\left( {{y_{j}},{z_{k}}}%
\right) }\right) {{\ln }_{q}}\frac{1}{{{1+\lambda p\left( {{y_{j}},{z_{k}}}%
\right) }}}}\right. \\ 
\left. {\,\,\,\,\,\,\,\,+\sum\limits_{i=1}^{n}{\left( {1+\lambda p\left( {{%
y_{j}},{z_{k}}}\right) \frac{p{\left( {{x_{i}},{y_{j}},{z_{k}}}\right) }}{p{%
\left( {{y_{j}},{z_{k}}}\right) }}}\right) }{{\ln }_{q}}\frac{1}{{1+\lambda
p\left( {{y_{j}},{z_{k}}}\right) \frac{p{\left( {{x_{i}},{y_{j}},{z_{k}}}%
\right) }}{p{\left( {{y_{j}},{z_{k}}}\right) }}}}}\right\} \\ 
\,\,\,\,\,\,+\frac{1}{{{\lambda ^{q}}}}\left\{ {-\left( {1+\lambda }\right) {%
{\ln }_{q}}\frac{1}{{1+\lambda }}+\sum\limits_{j=1}^{m}{\sum\limits_{k=1}^{l}%
{\left( {1+\lambda p\left( {{y_{j}},{z_{k}}}\right) }\right) {{\ln }_{q}}%
\frac{1}{{1+\lambda p\left( {{y_{j}},{z_{k}}}\right) }}}}}\right\} \\ 
=\frac{1}{{{\lambda ^{q}}}}\left\{ {-\left( {1+\lambda }\right) {{\ln }_{q}}%
\frac{1}{{1+\lambda }}+\sum\limits_{i=1}^{n}\sum\limits_{j=1}^{m}{%
\sum\limits_{k=1}^{l}{\left( {1+\lambda p\left( {{x_{i}},{y_{j}},{z_{k}}}%
\right) }\right) {{\ln }_{q}}\frac{1}{{1+\lambda p\left( {{x_{i}},{y_{j}},{%
z_{k}}}\right) }}}}}\right\} \\ 
={H_{\lambda ,q}}\left( {X,Y,Z}\right) .%
\end{array}%
\end{equation*}

\hfill \hbox{\rule{6pt}{6pt}}

\begin{The}
\label{the02_sec2.2} We assume $h(\lambda,q)=\lambda^{1-q}$. The strong
subadditivity for the Tsallis hypoentropies, 
\begin{equation}
H_{\lambda ,q}(X,Y,Z)+H_{\lambda ,q}(Z)\leq H_{\lambda ,q}(X,Z)+H_{\lambda
,q}(Y,Z),
\end{equation}%
holds for $1\leq q\leq 2$.
\end{The}

\textit{Proof:} This theorem is proven in a similar way as Theorem \ref%
{the01_sec2.2}. By the concavity of the function $Ln_{\lambda p(z_{k}),q}(x)$
in $x$, we have 
\begin{equation*}
\sum_{j=1}^{m}p(y_{j}|z_{k})Ln_{\lambda
p(z_{k}),q}(p(x_{i}|y_{j},z_{k}))\leq Ln_{\lambda p(z_{k}),q}\left(
\sum_{j=1}^{m}p(y_{j}|z_{k})p(x_{i}|y_{j},z_{k})\right) .
\end{equation*}%
Multiplying both sides by $p(z_{k})^{q}$ and summing over $i$ and $k$, we
have 
\begin{eqnarray}
&&\hspace*{-1cm}\sum_{j=1}^{m}\sum_{k=1}^{l}p(z_{k})^{q}p(y_{j}|z_{k})%
\sum_{i=1}^{n}Ln_{\lambda p(z_{k}),q}(p(x_{i}|y_{j},z_{k}))  \notag \\
&&\hspace*{1cm}\leq \sum_{k=1}^{l}p(z_{k})^{q}\sum_{i=1}^{n}Ln_{\lambda
p(z_{k}),q}(p(x_{i}|z_{k})),  \label{SSA_proof_ineq01}
\end{eqnarray}%
since $\sum_{j=1}^{m}p(y_{j}|z_{k})p(x_{i}|y_{j},z_{k})=p(x_{i}|z_{k})$. By $%
p(y_{j}|z_{k})^{q}\leq p(y_{j}|z_{k})$ for all $j$, $k$ and $1\leq q\leq 2$,
and by the nonnegativity of the function $Ln_{\lambda p(z_{k}),q}$, we have 
\begin{equation*}
p(y_{j}|z_{k})^{q}\sum_{i=1}^{n}Ln_{\lambda
p(z_{k}),q}(p(x_{i}|y_{j},z_{k}))\leq
p(y_{j}|z_{k})\sum_{i=1}^{n}Ln_{\lambda p(z_{k}),q}(p(x_{i}|y_{j},z_{k})).
\end{equation*}%
Multiplying both sides by $p(z_{k})^{q}$\ and summing over $j$ and $k$ in
the above inequality, we have 
\begin{eqnarray}
&&\hspace*{-1cm}\sum_{j=1}^{m}\sum_{k=1}^{l}p(z_{k})^{q}p(y_{j}|z_{k})^{q}%
\sum_{i=1}^{n}Ln_{\lambda p(z_{k}),q}(p(x_{i}|y_{j},z_{k}))  \notag \\
&&\hspace*{1cm}\leq
\sum_{j=1}^{m}\sum_{k=1}^{l}p(z_{k})^{q}p(y_{j}|z_{k})\sum_{i=1}^{n}Ln_{%
\lambda p(z_{k}),q}(p(x_{i}|y_{j},z_{k})).  \label{SSA_proof_ineq02}
\end{eqnarray}%
From the two inequalities (\ref{SSA_proof_ineq01}) and (\ref%
{SSA_proof_ineq02}) we have 
\begin{equation*}
\sum_{j=1}^{m}\sum_{k=1}^{l}p(z_{k})^{q}p(y_{j}|z_{k})^{q}\sum_{i=1}^{n}Ln_{%
\lambda p(z_{k}),q}(p(x_{i}|y_{j},z_{k}))\leq
\sum_{k=1}^{l}p(z_{k})^{q}\sum_{i=1}^{n}Ln_{\lambda
p(z_{k}),q}(p(x_{i}|z_{k})),
\end{equation*}%
which implies 
\begin{equation*}
\sum_{j=1}^{m}\sum_{k=1}^{l}p(y_{j},z_{k})^{q}\sum_{i=1}^{n}Ln_{\lambda
p(y_{j},z_{k}),q}(p(x_{i}|y_{j},z_{k}))\leq
\sum_{k=1}^{l}p(z_{k})^{q}\sum_{i=1}^{n}Ln_{\lambda
p(z_{k}),q}(p(x_{i}|z_{k})),
\end{equation*}%
since $p(y_{j},z_{k})\leq p(z_{k})$ (because of $%
\sum_{j=1}^{m}p(y_{j},z_{k})=p(z_{k})$) for all $j$ and $k$ and the function 
$Ln_{\lambda p(z_{k}),q}$ is monotonically increasing in $\lambda p(z_{k})>0$%
, when $1\leq q\leq 2$. Thus we have $H_{\lambda ,q}(X|Y,Z)\leq H_{\lambda
,q}(X|Z)$ which is equivalent to the inequality 
\begin{equation*}
H_{\lambda ,q}(X,Y,Z)-H_{\lambda ,q}(Y,Z)\leq H_{\lambda ,q}(X,Z)-H_{\lambda
,q}(Z)
\end{equation*}%
by Lemma \ref{lemma01_sec2.2} and Lemma \ref{lemma02_sec2.2}.

\hfill \hbox{\rule{6pt}{6pt}}

\begin{Rem}
Passing to the limit $\lambda \rightarrow \infty $ in Corollary \ref%
{cor01_sec2.2} and Theorem \ref{the02_sec2.2}, we recover the subadditivity
and the strong subadditivity \cite{Fu2006} for the Tsallis entropy: 
\begin{equation*}
T_{q}(X,Y)\leq T_{q}(X)+T_{q}(Y)\ \quad (q\geq 1)
\end{equation*}%
and 
\begin{equation*}
T_{q}(X,Y,Z)+T_{q}(Z)\leq T_{q}(X,Z)+T_{q}(Y,Z)\ \quad (q\geq 1).
\end{equation*}
\end{Rem}

Thanks to the subadditivities, we may define the Tsallis mutual
hypoentropies for $1\leq q\leq 2$ and $\lambda >0$. 

\begin{Def}
Let $1 \leq q \leq 2$ and $\lambda >0$. The Tsallis mutual hypoentropy is
defined by 
\begin{equation*}
I_{\lambda,q} (X;Y) \equiv H_{\lambda,q} (X) -H_{\lambda,q} (X|Y)
\end{equation*}
and the Tsallis conditional mutual hypoentropy is defined by 
\begin{equation*}
I_{\lambda,q} (X;Y|Z) \equiv H_{\lambda,q} (X|Z) -H_{\lambda,q} (X|Y,Z).
\end{equation*}
\end{Def}

From the chain rule given in Lemma \ref{lemma01_sec2.2}, we find that the
Tsallis mutual hypoentropy is symmetric, that is, 
\begin{eqnarray}
I_{\lambda ,q}(X;Y) &\equiv &H_{\lambda ,q}(X)-H_{\lambda ,q}(X|Y)  \notag \\
&=&H_{\lambda ,q}(X)+H_{\lambda ,q}(Y)-H_{\lambda ,q}(X,Y)  \notag \\
&=&H_{\lambda ,q}(Y)-H_{\lambda ,q}(Y|X)=I_{\lambda ,q}(Y;X).
\end{eqnarray}
In addition, we have 
\begin{equation}
0\leq I_{\lambda ,q}(X;Y)\leq \min \left\{ H_{\lambda ,q}(X),H_{\lambda
,q}(Y)\right\}
\end{equation}
from the subadditivity given in Theorem \ref{the01_sec2.2} and nonnegativity
of the Tsallis conditional hypoentropy. We also find $I_{\lambda
,q}(X;Y|Z)\geq 0$ from the strong subadditivity given in Theorem \ref%
{the02_sec2.2}.

Moreover we have the chain rule for the Tsallis mutual hypoentropies in the
following. 
\begin{eqnarray}
I_{\lambda ,q}(X;Y|Z) &=&H_{\lambda ,q}(X|Z)-H_{\lambda ,q}(X|Y,Z)  \notag \\
&=&H_{\lambda ,q}(X|Z)-H_{\lambda ,q}(X)+H_{\lambda ,q}(X)-H_{\lambda
,q}(X|Y,Z)  \notag \\
&=&-I_{\lambda ,q}(X;Z)+I_{\lambda ,q}(X;Y,Z).
\end{eqnarray}%
From the strong subadditivity, we have $H_{\lambda ,q}(X|Y,Z) \leq
H_{\lambda ,q}(X|Z)$, thus we have 
\begin{equation*}
I_{\lambda ,q}(X;Z)\leq I_{\lambda ,q}(X;Y,Z).
\end{equation*}
for $1 \leq q \leq 2$ and $\lambda >0$. 


\section{\protect\bigskip Jeffreys and Jensen-Shannon hypodivergences}

In what follows we indicate extensions of two known information measures.

\begin{Def}[\protect\cite{Dra2000},\protect\cite{Jef1946}]
The \emph{Jeffreys divergence} is defined by%
\begin{equation}
J(X||Y)\equiv D(X||Y)+D(Y||X)
\end{equation}%
and the \emph{Jensen-Shannon divergence} is defined as%
\begin{eqnarray}
JS(X||Y) &\equiv &\frac{1}{2}\left\{ D\left( X||\frac{X+Y}{2}\right)
+D\left( Y||\frac{X+Y}{2}\right) \right\}  \label{3} \\
&=&H\left( \frac{X+Y}{2}\right) -\frac{1}{2}\left( H\left( X\right) +H\left(
Y\right) \right) .
\end{eqnarray}
\end{Def}

The \emph{Jensen-Shannon divergence} was introduced in 1991 in \cite{Lin1991}%
, but its roots can be older, since one can see some analogous formulae used
in thermodynamics under the name \emph{entropy of mixing }\cite[p.598]%
{Tol1938}, for the study of gaseous, liquid or crystalline mixtures.

Jeffreys and Jensen-Shannon divergences have been extended to the context of
Tsallis theory in \cite{Fur2012}:

\begin{Def}
The Jeffreys-Tsallis divergence is 
\begin{equation}
J_{q}(X||Y)\equiv S_{q}(X||Y)+S_{q}(Y||X)
\end{equation}%
and the Jensen-Shannon-Tsallis divergence is 
\begin{equation}
JS_{q}(X||Y) \equiv \frac{1}{2}\left\{ S_{q}\left( X||\frac{X+Y}{2}\right)
+S_{q}\left( Y||\frac{X+Y}{2}\right) \right\}.
\end{equation}
\end{Def}

Note that 
\begin{equation*}
JS_{q}(X||Y) \neq T_{q}\left( \frac{X+Y}{2}\right) -\frac{1}{2}\left(
T_{q}\left( X\right) +T_{q}\left( Y\right) \right).
\end{equation*}
This expression was used in \cite{BenH2006} as \emph{Jensen-Tsallis
divergence}.

In accordance with the above definition, we define the directed\ Jeffreys
and Jensen-Shannon $q$- hypodivergence measures between two distributions
and emphasize the mathematical significance of our definitions.

\begin{Def}
The \emph{Jeffreys-Tsallis hypodivergence} is 
\begin{equation}
J_{\lambda ,q}(X||Y)\equiv D_{\lambda ,q}(X||Y)+D_{\lambda ,q}(Y||X)
\end{equation}%
and the \emph{Jensen-Shannon-Tsallis hypodivergence} is 
\begin{equation}
JS_{\lambda ,q}(X||Y)\equiv \frac{1}{2}\left\{ D_{\lambda ,q}\left( X||\frac{%
X+Y}{2}\right) +D_{\lambda ,q}\left( Y||\frac{X+Y}{2}\right) \right\} .
\end{equation}
\end{Def}

Here we point out that again one has 
\begin{eqnarray}
JS_{\lambda }(X||Y) &=&\frac{1}{2}K_{\lambda }\left( X|| \frac{X+Y}{2}%
\right) +\frac{1}{2}K_{\lambda }\left( Y|| \frac{X+Y}{2}\right) \\
&=&F_{\lambda }\left( \frac{X+Y}{2}\right) -\frac{1}{2}\left( F_{\lambda
}\left( X\right) +F_{\lambda }\left( Y\right) \right),
\end{eqnarray}%
where 
\begin{equation*}
JS_{\lambda }(X||Y)\equiv \lim_{q\rightarrow 1}JS_{\lambda ,q}(X||Y).
\end{equation*}

\begin{Lem}
\label{lemma2.32_sec2.3} The following inequality holds:%
\begin{equation*}
D_{\lambda ,q}\left( X||\frac{X+Y}{2}\right) \leq \frac{1}{2}D_{\lambda ,%
\frac{1+q}{2}}(X||Y)
\end{equation*}%
for $q\geq 0$ and $\lambda >0$.
\end{Lem}

\textit{Proof:} Using the inequality between the arithmetic and geometric
mean, one has

\begin{eqnarray}
D_{\lambda ,q}\left( X||\frac{X+Y}{2}\right) &=&-\frac{1}{\lambda }%
\sum_{i=1}^{n}\left( 1+\lambda p(x_i)\right) \ln _{q}\frac{\frac{\left(
1+\lambda p(x_i)\right) +\left( 1+\lambda p(y_i)\right) }{2}}{1+\lambda
p(x_i)} \\
&\leq &-\frac{1}{\lambda }\sum_{i=1}^{n}\left( 1+\lambda p(x_i)\right) \ln
_{q}\sqrt{\frac{1+\lambda p(y_i)}{1+\lambda p(x_i)}} \\
&=&-\frac{1}{2\lambda }\sum_{i=1}^{n}\left( 1+\lambda p(x_i)\right) \frac{%
\left( \frac{1+\lambda p(y_i)}{1+\lambda p(x_i)}\right) ^{1-\frac{1+q}{2}}-1%
}{1-\frac{1+q}{2}} \\
&=&\frac{1}{2}D_{\lambda ,\frac{1+q}{2}}(X||Y).
\end{eqnarray}

Thus the proof is completed.\hfill \hbox{\rule{6pt}{6pt}}

In the limit $\lambda \rightarrow \infty$, Lemma \ref{lemma2.32_sec2.3}
recovers Lemma 3.4 in \cite{Fur2012}.

\begin{Lem}[\protect\cite{Fur2012}]
\label{lemma_1}The function 
\begin{equation*}
f\left( x\right) =-\ln _{r}\frac{1+\exp _{q}x}{2}
\end{equation*}%
is concave\ for $0\leq r\leq q$.
\end{Lem}

The next two results of the present paper are stated in order to establish
the counterpart of Theorem 3.5 in \cite{Fur2012} for hypodivergences.

\begin{Prop}
It holds%
\begin{equation}
JS_{\lambda ,q}(X||Y)\leq \frac{1}{4}J_{\lambda ,\frac{1+q}{2}}(X||Y)
\end{equation}%
for $q\geq 0$ and $\lambda >0$.
\end{Prop}

\textit{Proof:} By the use of Lemma \ref{lemma2.32_sec2.3}, one has 
\begin{eqnarray}
2JS_{\lambda ,q}(X||Y) &=&D_{\lambda ,q}\left( X||\frac{X+Y}{2}\right)
+D_{\lambda ,q}\left( Y||\frac{X+Y}{2}\right) \\
&\leq &\frac{1}{2}D_{\lambda ,\frac{1+q}{2}}(X||Y)+\frac{1}{2}D_{\lambda ,%
\frac{1+q}{2}}(Y||X) \\
&=&\frac{1}{2}J_{\lambda ,\frac{1+q}{2}}(X||Y).
\end{eqnarray}%
This completes the proof.\hfill \hbox{\rule{6pt}{6pt}}

\begin{Prop}
\label{prop2.34_sec2.3} It holds that%
\begin{equation}
JS_{\lambda ,r}(X||Y)\leq -\frac{n+\lambda }{\lambda }\ln _{r}\frac{1+\exp
_{q}\left( -\frac{1}{2}\cdot \frac{\lambda }{n+\lambda }\cdot J_{\lambda
,q}(X||Y)\right) }{2}  \label{js_r}
\end{equation}%
for $0\leq r\leq q$ and $\lambda >0$.
\end{Prop}

\textit{Proof}: According to Lemma \ref{lemma_1},

\begin{equation*}
JS_{\lambda ,r}(X||Y)=%
\begin{array}{c}
-\frac{n+\lambda }{2\lambda }\left\{ \sum\limits_{i=1}^{n}\frac{1+\lambda
p(x_{i})}{n+\lambda }\ln _{r}\frac{1+\exp _{q}\ln _{q}\left( \frac{1+\lambda
p(y_{i})}{1+\lambda p(x_{i})}\right) }{2}+\sum\limits_{i=1}^{n}\frac{%
1+\lambda p(y_{i})}{n+\lambda }\ln _{r}\frac{1+\exp _{q}\ln _{q}\left( \frac{%
1+\lambda p(x_{i})}{1+\lambda p(y_{i})}\right) }{2}\right\}%
\end{array}%
\end{equation*}%
\begin{eqnarray}
&\leq &%
\begin{array}{c}
-\frac{n+\lambda }{2\lambda }\left\{ \ln _{r}\frac{1+\exp
_{q}\sum\limits_{i=1}^{n}\frac{1+\lambda p(x_{i})}{n+\lambda }\ln _{q}\left( 
\frac{1+\lambda p(y_{i})}{1+\lambda p(x_{i})}\right) }{2}+\ln _{r}\frac{%
1+\exp _{q}\sum\limits_{i=1}^{n}\frac{1+\lambda p(y_{i})}{n+\lambda }\ln
_{q}\left( \frac{1+\lambda p(x_{i})}{1+\lambda p(y_{i})}\right) }{2}\right\}%
\end{array}
\notag \\
&=&%
\begin{array}{c}
-\frac{n+\lambda }{2\lambda }\left\{ \ln _{r}\frac{1+\exp _{q}\left( -\frac{%
\lambda }{n+\lambda }D_{\lambda ,q}(X||Y)\right) }{2}+\ln _{r}\frac{1+\exp
_{q}\left( -\frac{\lambda }{n+\lambda }D_{\lambda ,q}(Y||X)\right) }{2}%
\right\}.%
\end{array}%
\end{eqnarray}%
Then%
\begin{eqnarray}
JS_{\lambda ,r}(X||Y) &\leq &-\frac{n+\lambda }{\lambda }\ln _{r}\frac{%
1+\exp _{q}-\frac{\lambda }{n+\lambda }\left( \frac{D_{\lambda
,q}(X||Y)+D_{\lambda ,q}(Y||X)}{2}\right) }{2}  \notag \\
&=&-\frac{n+\lambda }{\lambda }\ln _{r}\frac{1+\exp _{q}\left( -\frac{1}{2}%
\cdot \frac{\lambda }{n+\lambda }\cdot J_{\lambda ,q}(X||Y)\right) }{2}.
\end{eqnarray}%
Thus the proof is completed.\hfill \hbox{\rule{6pt}{6pt}}

We further define the dual symmetric hypodivergences.

\begin{Def}
The dual symmetric Jeffreys-Tsallis hypodivergence is defined by 
\begin{equation*}
J_{\lambda ,q}^{(ds)}(X||Y)\equiv D_{\lambda ,q}(X||Y)+D_{\lambda ,2-q}(Y||X)
\end{equation*}%
and the dual symmetric Jensen-Shannon-Tsallis hypodivergence is defined by 
\begin{equation*}
JS_{\lambda ,q}^{(ds)}(X||Y)\equiv \frac{1}{2}\left\{ D_{\lambda ,q}\left(X||%
\frac{X+Y}{2}\right)+D_{\lambda ,2-q}\left(Y||\frac{X+Y}{2}\right)\right\} .
\end{equation*}
\end{Def}

Using Lemma \ref{lemma2.32_sec2.3}, we have the following inequality.

\begin{Prop}
It holds 
\begin{equation*}
JS_{\lambda,q}^{(ds)}(X||Y) \leq \frac{1}{4}J_{\lambda,\frac{1+q}{2}%
}^{(ds)}(X||Y)
\end{equation*}
for $0 \leq q \leq 2$ and $\lambda >0$.
\end{Prop}

In addition, we have the following inequality.

\begin{Prop}
It holds 
\begin{equation*}
JS_{\lambda,q}^{(ds)}(X||Y) \leq -\frac{n+\lambda}{\lambda} \ln_r \frac{%
1+\exp_q\left(-\frac{\lambda}{2(n+\lambda)} J_{\lambda,q} (X||Y)\right)}{2}.
\end{equation*}
for $1<r\leq 2$, $r \leq q$ and $\lambda >0$.
\end{Prop}

\textit{Proof:} The proof can be done by similar calculations with
Proposition \ref{prop2.34_sec2.3}, applying the facts (see Lemma 3.9 and
3.10 in \cite{Fur2012}) that $\exp _{q}(x)$ is a monotonically increasing
function in $q$ for $x\geq 0$, and the inequality $-\ln _{2-r}x\leq -\ln
_{r}x$ holds for $1<r\leq 2$ and $x>0.$

\hfill \hbox{\rule{6pt}{6pt}}

\section{Concluding remarks}

In this paper, we introduced the Tsallis hypoentropy $H_{\lambda ,q}(X)$ and
studied some properties of $H_{\lambda ,q}(X)$. We named $H_{\lambda ,q}(X)$ 
\emph{Tsallis hypoentropy} because of the relation $H_{\lambda ,q}(X)\leq
T_{q}(X)$ which follows from the monotonicity given in Proposition \ref%
{Tsallis_hypoentropy_monotonicity_ineq01} and Lemma \ref{lemma_monotonicity}
for the case $h(\lambda ,q)=(1+\lambda )^{1-q}$ and the case $h(\lambda
,q)=\lambda ^{1-q}$, respectively (this relation can be also proven
directly). In this naming we follow Ferreri, as he has termed $F_{\lambda
}(X)$ \emph{hypoentropy} due to the relation $F_{\lambda }(X)\leq H(X)$.

The monotonicity of the hypoentropy and the Tsallis hypoentropy for $\lambda
>0$, indeed, is an interesting feature. It may be remarkable to examine the
monotonicity of the Tsallis entropy for the parameter $q\geq 0$. We find
that the Tsallis entropy $T_{q}(X)$ is monotonically decreasing with respect
to $q\geq 0$. Indeed, we find $\frac{dT_{q}(X)}{dq}=\sum_{j=1}^{n}\frac{%
p_{j}^{q}v_{q}(p_{j})}{(1-q)^{2}}$, where $v_{q}(x)\equiv
1-x^{1-q}+(1-q)\log x$ $(0\leq x\leq 1)$. Since $x^{q}v_{q}(x)=0$ for $x=0$
and $q>0$, we prove $v_{q}(x)\leq 0$ for $0<x\leq 1$. We find $\frac{%
dv_{q}(x)}{dx}=\frac{(1-q)(1-x^{1-q})}{x}\geq 0$ when $0<x\leq 1$, thus we
have $v_{q}(x)\leq v_{q}(1)=0$ which implies $\frac{dT_{q}(X)}{dq}\leq 0$.
This monotonicity implies the relations $H(X)\leq T_{q}(X)$ for $0\leq q<1$
and $T_{q}(X)\leq H(X)$ for $q>1$. (These relations are also proven by the
inequalities $\log \frac{1}{x}\leq \ln _{q}\frac{1}{x}$ for $0\leq q<1$, $%
x>0 $ and $\log \frac{1}{x}\geq \ln _{q}\frac{1}{x}$ for $q>1$, $x>0$.)

As another important results, we also gave the chain rules, subadditivity and
the strong subadditivity of the Tsallis hypoentropies in the case of $%
h(\lambda ,q)=\lambda ^{1-q}$. For the case of $h(\lambda ,q)=(1+\lambda
)^{1-q}$, we can prove $H_{\lambda ,q}(Y|X)\leq H_{\lambda ,q}(X)$ and $%
H_{\lambda ,q}(X|Y,Z)\leq H_{\lambda ,q}(X|Z)$ for $1\leq q\leq 2$ in a
similar way to the proofs of Theorem \ref{the01_sec2.2} and \ref%
{the02_sec2.2}, since the function $Sn_{\lambda ,q}(x)$ defined in the proof
of Proposition \ref{Tsallis_hypoentropy_monotonicity_ineq01} is also
nonnegative, monotone increasing and concave in $x\in \lbrack 0,1]$ and we
have $H_{\lambda p(x_{i}),q}(Y|x_{i})=\sum_{j=1}^{m}Sn_{\lambda
p(x_{i}),q}(p(y_{j}|x_{i}))$ for all fixed $x_i$. However we cannot obtain
the inequalities 
\begin{equation*}
H_{\lambda ,q}(X,Y)\leq H_{\lambda ,q}(X)+H_{\lambda ,q}(Y)\ \quad (1\leq
q\leq 2),
\end{equation*}%
\begin{equation*}
H_{\lambda ,q}(X,Y,Z)+H_{\lambda ,q}(Z)\leq H_{\lambda ,q}(X,Z)+H_{\lambda
,q}(Y,Z)\ \quad (1\leq q\leq 2)
\end{equation*}%
for $h(\lambda ,q)=(1+\lambda )^{1-q}$, because the similar proof for the
chain rules does not work well in the case $h(\lambda ,q)=(1+\lambda )^{1-q}$%
.

\section*{Acknowledgement}
The author (S.F.) was partially supported by JSPS KAKENHI Grant Number 24540146.

\end{document}